\documentclass[11pt]{article}
\newif\iffigs\figstrue

\usepackage{epsfig,latexsym}
\usepackage{amsmath}
\usepackage{verbatim}
\usepackage{mathrsfs}
\usepackage{amssymb}

\DeclareMathAlphabet{\mathpzc}{OT1}{pzc}{m}{it}

 \csname
@addtoreset\endcsname{equation}{section}

\def\gz0{\gamma^{0}}

\def\be{\begin{equation}}
\def\ee{\end{equation}}
\def\bea{\begin{eqnarray}}
\def\eea{\end{eqnarray}}
\def\ba{\begin{array}}
\def\ea{\end{array}}
\def\bec{\begin{center}}
\def\ec{\end{center}}
\def\ba{\begin{align}}
\def\ena{\end{align}}

\def\12{\frac{1}{2}}

\newcommand{\I}{\mathrm{Im}}

\begin{document}

\begin{flushright}
CERN-PH-TH/2014-040\\
CPHT-RR010.0314 \\
{\today}
\end{flushright}

\vspace{10pt}

\begin{center}


{\Large\sc The Volkov--Akulov--Starobinsky Supergravity}\\


\vspace{25pt}
{\sc I.~Antoniadis${}^{\; a}$\footnote{On leave of absence from CPhT, \'Ecole Polyt\'echnique, F-91128 Palaiseau Cedex, France.}, E.~Dudas${}^{\; b}$, S.~Ferrara${}^{\; a,c,d}$ and A.~Sagnotti$^{\; e}$}\\[15pt]

{${}^a$\sl\small Department of Physics, CERN Theory Division\\
CH - 1211 Geneva 23, Switzerland \\ }e-mail: {\small \it
Ignatios.Antoniadis@cern.ch, Sergio.Ferrara@cern.ch}\vspace{6pt}

{${}^b$\sl\small Centre de Physique Th\'eorique,
\'Ecole Polyt\'echnique\\
F-91128 Palaiseau \ FRANCE\\}e-mail: {\small \it
dudas@cpht.polytechnique.fr}\vspace{6pt}

{${}^c$\sl\small INFN - Laboratori Nazionali di Frascati \\
Via Enrico Fermi 40, I-00044 Frascati, Italy}\vspace{6pt}

{${}^d$\sl\small Department of Physics and Astronomy,
University of California \\ Los Angeles, CA 90095-1547, USA}
\vspace{6pt}

{${}^e$\sl\small
Scuola Normale Superiore and INFN\\
Piazza dei Cavalieri \ 7\\I-56126 Pisa \ ITALY \\}
e-mail: {\small \it sagnotti@sns.it}
\vspace{12pt}

\vspace{40pt} {\sc\large Abstract}
\end{center}

We construct a supergravity model whose scalar degrees of freedom arise from a chiral superfield and are solely a scalaron and an axion that is very heavy during the inflationary phase. The model includes a second chiral superfield $X$, which is subject however to the constraint $X^2=0$ so that it describes only a Volkov--Akulov goldstino and an auxiliary field. We also construct the dual higher--derivative model, which rests on a chiral scalar curvature superfield ${\cal R}$ subject to the constraint ${\cal R}^2=0$, where the goldstino dual arises from the gauge--invariant gravitino field strength as $\gamma^{mn} {\cal D}_m \psi_n$. The final bosonic action is an $R+R^2$ theory involving an axial vector $A_m$ that only propagates a physical pseudoscalar mode.

\noindent

\setcounter{page}{1}

\pagebreak

\newpage
\section{\sc  Starobinsky models of inflation in Supergravity}\label{sec:staro}
It was recently shown how to embed in Supergravity \cite{SUGRA} a class of models \cite{fkvp} including the Starobinsky potential \cite{starobinsky}, which affords an excellent agreement with recent PLANCK data \cite{planck} for an inflationary epoch of about $60$ $e$--folds. However, the models based on the ``old minimal'' supergravity rest on a pair of chiral superfields (see also \cite{kl, starobinsky_sugra,kallosh} for closely related work), and thus involve three scalar fields in addition to the inflaton.

The construction reflects the Starobinsky duality \cite{starobinsky, whitt} between an $R+R^2$ action and a special scalar--gravity system, encompassed by the master action
\be
{\cal S}
\ = \ \int d^{\,4} x \sqrt{-g} \left[ \ \frac{1}{2} \ R \ {+} \ \chi \big(R \,-\, \varphi\big) \ + \ \alpha\, \varphi^{\,2} \right] \, , \label{staro0}
\ee
which reduces to
\be
{\cal S}_1  \ = \ \int d^{\,4} x \sqrt{-g} \left[ \ {{{1\over 2} (1+\,2\,\chi )}} R \ - \ {1\over 4\alpha}\ \chi^2\right] \label{staro0dual}
\ee
upon integration over $\varphi$. On the other hand, $\chi$ enters eq.~(\ref{staro0}) as a Lagrange multiplier imposing the constraint $\varphi=R$, and enforcing it leads to the dual form
\be
{\cal S}_2 \ = \ \int d^{\,4} x \sqrt{-g} \left[ \ \frac{1}{2} \ R \ + \ \alpha\, R^{\,2}\right] \ .
\ee

The embedding of this model in (higher--derivative) supergravity takes the form \cite{cecotti,fkvp}
\begin{equation}
{\cal L} = \left[- \, S_0\, {\overline S}_0 \ +\ h \left(\frac{\cal R}{S_0},\frac{\cal {\overline R}}{{\overline S}_0} \right) S_0 \, {\overline S}_0  \right]_D + \left[W\left(\frac{\cal R}{S_0}\right) S_0^3 \right]_F \ , \label{staro1}
\end{equation}
where here and in the following subscripts identify $D$ and $F$ superspace densities, $S_0$ is the chiral compensator field and $ {\cal R}$ is the chiral scalar curvature superfield, defined via the curved chiral projector $\Sigma$ as
\be
{\cal R} \ = \ \frac{\Sigma ({\overline S}_0)}{S_0} \ .
\ee
In eq.~(\ref{staro1}) $h$ is a real function of the chiral superfield ${\cal R}/{S_0}$ and its conjugate ${\overline{\cal R}}/{\overline{S}_0}$ that contains an ${\cal R}{\cal {\overline R}}$ term and $W$ is a chiral function. The presence of the ${\cal R}{\cal {\overline R}}$ term in $h$ brings about an $R^2$ term in components.

In detail, the components of the chiral superfield ${\cal R}$ are \cite{SUGRA,fv,fkvp}
\begin{equation}
{\cal R} = \left( {\overline u} \equiv S \,+\, i P\, , \ \gamma^{mn} {\cal D}_m \psi_n \, ,\ - \ \frac{1}{2}\ R
\,-\, \frac{1}{3} \ A_m^2 \,+\, i \,{\cal D}^m A_m \,- \,\frac{1}{3} \ u \, {\overline u}  \right)  \ , \label{staro01}
\end{equation}
where $u$ and $A_m$ are the ``old minimal'' auxiliary fields of $N=1$ supergravity and $\psi_n$ is the gravitino field. The action (\ref{staro1}) can be recast in a two--derivative dual form proceeding along the lines of eqs.~(\ref{staro0}) and (\ref{staro0dual}) and making use of a pair of chiral multiplets $\Lambda$ and $C$ as in \cite{cecotti,fkvp}, so that
\begin{equation}
{\cal L} = \bigg[- \ S_0 \, {\overline S}_0 \, + \, h(C,{\overline C})\, S_0 \, {\overline S}_0 \bigg]_D \ + \
\left[  \left\{ \Lambda \left(C - \frac{\cal R}{S_0}\right) \, + \, W(C) \right\} S_0^3 \right]_F \ . \label{staro2}
\end{equation}

Eliminating $\Lambda$ in (\ref{staro2}) yields the constraint ${\cal R} = S_0 C$, and one recovers in this fashion the original Lagrangian (\ref{staro1}). Notice, to this end, that letting
\be
W(C) \ = \ C \, g(C)\ + \ W_0 \ , \label{staro25}
\ee
on account of the identity \cite{SUGRA,fkvp}
\begin{equation}
\bigg[ f(\Lambda)\, {\cal R}\, S_0^2 \bigg]_F \ + \ {\rm h.c.} \ = \ \bigg[ \big(f(\Lambda) \, + \, {\overline f} ({\overline \Lambda})\big)
S_0 \, {\overline S}_0  \bigg]_D  \ + \ {\rm tot. \ deriv.} \ , \label{staro3}
\end{equation}
which holds for any chiral superfield $\Lambda$ and for any function $f$, $g(C)$ can be shifted away redefining $\Lambda$ into another chiral superfield $\Lambda'$ \cite{fkvp}, so that the Lagrangian can be cast in the form
\begin{equation}
{\cal L}_{\rm dual} = \bigg[ \big[-1 - \Lambda' - {\overline \Lambda}{\,'} \, + \, h(C,{\overline C})\big] \, S_0 \, {\overline S}_0
\bigg]_D + \bigg[ (W_0 \, + \, C\, \Lambda')\ + \ {\rm h.c. } \bigg]_F  \ . \label{staro4}
\end{equation}
Letting
\be
\Lambda' \ = \ T \ - \ \frac{1}{2} \ ,
\ee
one is then finally led to a standard $N=1$ supergravity with a K\"ahler potential $K$ and a superpotential $W$ given by
\begin{equation}
K \ =\  - \, 3 \, \ln \left[T \, + \, {\overline T} \, -\,  h (C,{\overline C})\right] \ , \qquad W \, = \, C \, \left(T\, -\, \frac{1}{2}\right) \ + \ W_0 \ . \label{staro5}
\end{equation}
The case in which $h(C,{\overline C})$ is a pure K\"ahler transformation of the form $h(C) + {\overline h} ({\overline C})$, so that $C$ is not dynamical, was considered in \cite{ketov}, where it was referred to as
$f(R)$ supergravity, but this class of models does not reproduce the $R+R^2$ bosonic terms and the Starobinsky potential \cite{fkvp,kehagias,kt}.

The Lagrangian (\ref{staro5}) contains Starobinsky's inflaton $\phi$, which is related to $T$ according to $Re(T) = \exp\big(\sqrt{{2}/{3}}\ \phi\big)$, and setting to zero the other three fields one can recover exactly, for $W_0=0$, the scalar potential of the original Starobinsky model. However, it was shown in \cite{kl} that for a minimal choice $h (C,{\overline C}) = C {\overline C}$  the complex scalar direction $C$ is \emph{unstable} during the inflationary phase. Non--minimal (and therefore non--universal) K\"ahler functions are thus needed to arrive at a satisfactory model, and in particular Kallosh and Linde \cite{kl} showed that
\begin{equation}
h (C,{\overline C}) \ =\  C \, {\overline C} \ - \ \zeta \, (C \, {\overline C})^2 \  \label{staro6}
\end{equation}
can stabilize the $C$ direction for sufficiently large positive values of $\zeta $.

Although for $W_0=0$ the model admits a supersymmetric ground state, supersymmetry is broken during the inflationary phase, and $C$ plays the role of a goldstino superfield driving the breaking of supersymmetry \cite{fkvp,kl,kallosh}. In what follows we shall explain how a minimal and universal embedding of the Starobinsky potential in Supergravity emerges once the ordinary chiral superfield $C$ is replaced with a chiral superfield $X$ satisfying, as in \cite{rocek,lindstrom,casalbuoni,ks}, the constraint $X^2=0$.
\vskip 15pt
\section{\sc  The Volkov--Akulov Lagrangian}\label{sec:va}
It has been known for some time that the Volkov--Akulov Lagrangian \cite{va} can be recast in a manifestly supersymmetric form introducing a chiral superfield $X$ that satisfies identically the constraint \cite{rocek,lindstrom,casalbuoni,ks}
\be
X^2=0 \label{va0} \ .
\ee
This eliminates the scalar component of $X$ in favor of a goldstino bilinear, so that in two--component notation
\begin{equation}
X \ = \ \frac{GG}{2 \, F_X} \ + \ \sqrt{2}\, \theta\, G \ +\  \theta^2 F_X \ , \label{va1}
\end{equation}
and the complete Volkov--Akulov Lagrangian is then
\begin{equation}
{\cal L}_{VA} \ = \ \bigg[ X\, \overline{X} \bigg]_D \ + \ \bigg[f X \,+\, h.c. \bigg]_F \ , \label{va2}
\end{equation}
where the subscripts denote again $D$ and $F$ superspace densities. Notice that
supersymmetry can be realized \emph{off--shell}, as emphasized in \cite{ks,adgt}, insofar as $F_X$ is not replaced by the solution of its algebraic equation of motion. In Supergravity, the off--shell couplings of the goldstino to the gravity multiplet can be found in a similar way, but taking into account the constraint \eqref{va0} the most general couplings of $X$ to Supergravity rest on a K\"ahler potential $K$ and a superpotential $W$ of the form
\begin{equation}
K \ = \ - \ 3\, \log\left(1 \,-\, X \, \overline{X}\right) \ \equiv \ 3\, X \, \overline{X} \ , \qquad W \ = \ f\, X \ + \ W_0 \ , \label{va3}
\end{equation}
since terms linear in $X$ or $\overline{X}$ can be reabsorbed in $W$, and as a result \cite{cfgvp}
\be
V \ = \ \frac{1}{3}\ |f |^2 \ - \ 3\, |W_0|^2 \ , \qquad m_{3/2}^2 \ = \ |W_0|^2 \ .
\ee
The supergravity Lagrangian resulting from eq.~\eqref{va3} does encode the proper goldstino couplings, and in particular in two--component notation the fermionic mass terms read
\begin{equation}
{\cal L}_{mass} \ = \ -\ m_{3/2} \left( \psi_m \ + \ \frac{i}{\sqrt{6}} \ \sigma_m \,{\overline G} \right)
\sigma^{mn} \left( \psi_n \ + \ \frac{i}{\sqrt{6}} \ \sigma_n \, {\overline G} \right)\ + \ {\rm h.c.} \ . \label{va4}
\end{equation}
\vskip 15pt
\section{\sc  The minimal Starobinsky Lagrangian }\label{sec:vas}

The usual embedding of the Starobinsky Lagrangian in Supergravity rests, in the ``old minimal'' two--derivative formulation, on the gravitational supermultiplet coupled to a pair of additional chiral multiplets. As we have anticipated, the corresponding action is not unique, and
non--minimal terms are actually needed \cite{kl} to stabilize the scalar fields during the inflationary phase.

A minimal universal model obtains if Supergravity is coupled to the \emph{constrained} goldstino multiplet $X$ described in the preceding section and to a chiral multiplet $T$ containing the inflaton\footnote{The constrained superfield $X$ was previously considered, in a different context also related to inflationary models, in \cite{luis}.}. Taking into account the constraint of eq.~\eqref{va0}, in this \emph{off--shell} formulation the Lagrangian is determined by
\be
K \ = \ - \ 3 \ \ln \big[T + {\overline T} - X\, \overline{X} \big] \ , \qquad W \ = \ M \, X \, T \ + \ f \, X \ + \  W_0 \ , \label{vas1}
\ee
where $M^2=\frac{3}{4\alpha}$ from the comparison with eq.~(\ref{staro0}),
while the corresponding scalar potential \cite{cfgvp} is simply
\begin{equation}
V \ = \ \frac{|M T \,+ \, f|^2}{3 \, ( T \, + \, {\overline T})^2} \ , \label{vas2}
\end{equation}
since the scalar component of $X$ is not a dynamical field but a goldstino bilinear. Let us stress, however, that $X$ contributes to the scalar potential via its derivatives, since
\be
F_{\overline{X}} \ = \ { e^{\frac{K}{2}}} \left( K_{X\,\overline{X}} \right)^{\,-\,1} \ {\overline{W}}_{\overline{X}} \ ,
\ee
includes a bosonic contribution although it contains no elementary scalar field.

The form of (\ref{vas2}) reflects the no--scale structure \cite{noscale} of the $T$ kinetic term and its coupling to the goldstino superfield $X$, and in particular the constant superpotential $W_0$ does not enter $V$ while it determines the gravitino mass. The complete
bosonic Lagrangian is
\begin{equation}
{\cal L} \ = \ \frac{R}{2} \ - \ \frac{3}{(T \,+ \, {\overline T})^2} \ |\partial \,T|^2\  - \ \frac{|M \,T\, + \, f|^2}{3 ( T \,+ \,{\overline T})^2}   \ , \label{vas3}
\end{equation}
and letting
\be
T \ = \ e^{\,\phi{\sqrt \frac{2}{3}}} \ + \ i \, a \, \sqrt{\frac{2}{3}} \ ,
\ee
it finally becomes
\begin{equation}
{\cal L} \ = \ \frac{R}{2} \ - \ \frac{1}{2}\ (\partial \phi)^2 \ - \ \frac{1}{2} \ e^{\,-\,2\, \phi \,{\sqrt \frac{2}{3}}}\
(\partial a)^2 \ - \ \frac{1}{12} \left(M \ + \ f \, e^{\,-\,{\sqrt \frac{2}{3}} \ \phi}\right)^2 - \frac{M^2}{18} \ e^{- 2\,\phi
{\sqrt \frac{2}{3}}} \ a^2 \ . \label{vas4}
\end{equation}
If $M f < 0$, after a shift of $\phi$ and a rescaling of the axion $a$, one can bring the Lagrangian to the form
 \begin{equation}
{\cal L} \ = \ \frac{R}{2} \ - \ \frac{1}{2}\ (\partial \phi)^2 \ - \ \frac{1}{2} \ e^{\,-\,2\, \phi \,{\sqrt \frac{2}{3}}}\
(\partial a)^2 \ - \ \frac{M^{\,2}}{12} \left(1 \ - \ e^{\,-\,{\sqrt \frac{2}{3}} \ \phi}\right)^2 - \frac{M^2}{18} \ e^{- 2\,\phi
{\sqrt \frac{2}{3}}} \ a^2 \ . \label{vas5}
\end{equation}
This is precisely a minimal Starobinsky Lagrangian where $\phi$ is accompanied by an axion field $a$ that is however much heavier during the inflationary phase where the vacuum values $\phi_0$ are large and positive, so that
\begin{equation}
m_{\phi}^2 \ \simeq \ \frac{M^2}{9} \ e^{\,-\,2\, \phi_0 {\sqrt \frac{2}{3}}} \ << \
m_{a}^2 \ \equiv \ \frac{M^2}{9}  \ . \label{vas6}
\end{equation}
\vskip 15pt
\section{\sc  Dual gravitational formulation} \label{sec:dual}
In the conformal compensator formalism \cite{SUGRA}, the Lagrangian of the previous section reads
\begin{equation}
{\cal L} \ = \ - \ \bigg[ \big(T \,+ \,{\overline T} \,- \, |X|^2\big) S_0 \, {\overline S}_0 \bigg]_D \ +\
\bigg[ \big(M X T \,+ \,f X \,+\, W_0\big) S_0^3 \, + \, {\rm h.c} \bigg]_F
 \ , \label{dual1}
 \end{equation}
and can be recast in the form
\begin{equation}
{\cal L} \ =  \ \bigg[ |X|^2 \,S_0 \, {\overline S}_0 \bigg]_D \ +\
\left[ \left(T\big(-\,\frac{\cal R}{S_0}\,+ \, M \, X\big)  \,+ \,f \,X \,+ \,W_0 \right) S_0^3 \ + \ {\rm h.c} \right]_F
 \label{dual2}
 \end{equation}
resorting to the identity
\be
\bigg[ (T + {\overline T}) S_0 \, {\overline S}_0\bigg]_D \ = \ \bigg[T \, {\cal R} \, S_0^2\bigg]_F \ + \ {\rm h.c.} \ ,
\ee
where $ {\cal R}$ is the chiral supergravity multiplet. Notice that $T$ enters eq.~\eqref{dual2} as a Lagrange multiplier, whose equation of motion is the constraint
\begin{equation}
X \ =  \ \frac{1}{M}\frac{\cal R}{S_0}  \ , \label{dual3}
\end{equation}
and as a result the constraint $X^2 =0$ of eq.~\eqref{va0} translates into a similar constraint on ${\cal R}$:
\begin{equation}
{\cal R}^2 = 0  \ . \label{dual4}
\end{equation}

Conversely, one can start from the dual gravitational theory
\begin{equation}
{\cal L} \ = \ - \ \left[ S_0 \, {\overline S}_0 \, - \, \frac{{\cal R} {\cal {\overline R}}}{M^2} \right]_D
\ + \ \left[ W_0\, + \, \xi \, \frac{{\cal R}}{S_0} \ S_0^3 \, + \, \sigma \, {\cal R}^2 \, S_0 \right]_F   \ , \label{dual5}
\end{equation}
and in this form the nonlinear constraint (\ref{dual4}) originates from the field equation of $\sigma$. One can also use the identity
\begin{equation}
\bigg[ \sigma \, {\cal R}^2\, S_0 \, + \, h.c.  \bigg]_F \ = \ \left[ \left(\sigma \, \frac{\cal R}{S_0}\, +\,
{\overline \sigma} \, \frac{\cal {\overline R}}{{\overline S}_0}\right) S_0 \, {\overline S}_0 \right]_D \ + \ {\rm tot. \ deriv.} \ , \label{dual05}
\end{equation}
which is a particular case of  (\ref{staro3}), and introducing two Lagrangian chiral superfields multipliers $T$ and $C$ according to
\begin{eqnarray}
{\cal L}_{\rm dual} &=& - \left[ \bigg(1 \, {-} \, \sigma\, C\,   {-} \, {\overline \sigma}\, {\overline C} -
\frac{ C \,{\overline C}}{M^2} \bigg) S_0 \,{\overline S}_0 \right]_D +\,
\left[ \left( - T \big(\frac{\cal R}{S_0}\,-\,C\big) + W_0 + \xi C \right) S_0^3 + {\rm h.c. } \right]_F
\nonumber \\
&=& - \, \left[ \bigg(1 \,+ \,T \,+\, {\overline T} \, {-} \, \sigma \,C \, {-} \, {\overline \sigma}\, {\overline C} \,-\,
\frac{C \, {\overline C}}{M^2}\bigg) S_0 \, {\overline S}_0 \right]_D  +\,
\left[ {\widetilde W} (\sigma,T,C) \, S_0^3 \,+\, {\rm h.c. } \right]_F   , \label{dual051}
\end{eqnarray}
where
\begin{equation}
{\widetilde W} (T,C) \ = \ W_0 \ + \ (T \, + \, \xi ) C   \ , \label{dual052}
\end{equation}
a final shift $T \ \to \ T \  {+}  \ \sigma \, C \ - \ \frac{1}{2}$ and the replacement of
$\xi - \frac{1}{2}$ with $f$ turn this expression into
\begin{equation}
{\widetilde W} (\sigma,T,C) \ = \ T \, C \ + \ \sigma \, C^2 \ + \ W_0\  +\  f \, C  \ . \label{dual053}
\end{equation}
Notice that the $\sigma$ field equation enforces the constraint $X^2=0$, and finally letting $C/M = X $ and recaling $f$ to $f/M$ yields the Lagrangian
\be
{\cal L} \ = \  - \ \bigg[ (T \, + \, {\overline T} \, - \, X \, {\overline X}) \, S_0 \, {\overline S}_0 \bigg]_D \ + \ \bigg[ W (T,X) \, S_0^3 \, +\,
{\rm h.c.} \bigg]_F  \ , \label{dual054}
\ee
where
\begin{equation}
W (T,X) \  = \ W_0 \ + \ ( M \,T \ + \ f ) X    \ . \label{dual055}
\end{equation}
This is precisely the dual action coupled to a goldstino multiplet (\ref{vas1}), which completes our proof of the duality with the higher--derivative supergravity form of the Starobinsky model (\ref{dual5}).

In order to write explicitly the bosonic gravitational action, one introduces the Jordan scalar
\begin{equation}
e^{\, \phi{\sqrt \frac{2}{3}}} \ = \ 1 \ + \ 2 \, \chi  \ , \label{dual6}
\end{equation}
in terms of which the Lagrangian (\ref{vas5}) becomes
\begin{equation}
{\cal L} \ = \ \frac{R}{2} \ - \ \frac{3}{(1 \,+ \, 2\, \chi)^2} \ \left[(\partial \chi)^2 \, +\, \frac{1}{6}\ (\partial a)^2\right]\
- \ \frac{M^2}{3} \ \frac{\chi^2\, +\, \frac{a^2}{6}}{(1 \, +\, 2\, \chi)^2}  \ . \label{dual7}
\end{equation}
The transition to the Jordan frame is effected by the Weyl rescaling
\begin{equation}
g \ \rightarrow \ \left( 1 \ + \ 2 \, \chi \right) g \ ,  \label{dual8}
\end{equation}
and the resulting Lagrangian is
\begin{equation}
{\cal L} \ = \ \frac{1}{2} \ (1 \, + \, 2 \, \chi)\  R  \ - \ \frac{1}{2} \ \frac{(\partial a)^2}{1 \,+ \, 2 \, \chi} \ - \ \frac{M^2}{3} \ \left(\chi^2\, +\, \frac{a^2}{6}\right)   \ . \label{dual9}
\end{equation}

In the ``old minimal'' supergravity formulation the axion should be traded for the longitudinal mode ${\cal D} \cdot A$ of the pseudovector auxiliary field $A_m$, and this last step brings about an interesting general link that can be deduced from the master Lagrangian
\begin{equation}
{\cal L}_a \ = \ -\  \frac{M^2}{18} \ a^2 \ + \ A^m \, \partial_m \,a \ +\  \frac{1}{2} \ (1 \,+\, 2\, \chi) A_m^2  \ . \label{dual10}
\end{equation}
Varying ${\cal L}_a $ with respect to $A_m$ yields indeed
\begin{equation}
A_m \ = \ - \ \frac{\partial_m \,a}{1 \,+ \, 2 \, \chi} \ , \qquad {\cal L}_a \ = \ - \ \frac{M^2}{18} \ a^2 \ - \ \frac{1}{2}
\ \frac{(\partial a)^2}{1 \,+\, 2 \, \chi}  \ , \label{dual11}
\end{equation}
while varying ${\cal L}_a $ with respect to $a$ yields
\begin{equation}
a \ =\  -\ \frac{9}{M^2} \ {\cal D} \cdot A \ , \qquad {\cal L}_a \ = \ \frac{9}{2 M^2} \ ({\cal D} \cdot A)^2
\ + \ \frac{1}{2} \ (1\, + 2\, \chi) \ A_m^2 \ . \label{dual12}
\end{equation}
Using these results in (\ref{dual9}) one finds
\begin{equation}
{\cal L} \ = \ \frac{1}{2} \ (1 \, +\, 2 \, \chi) (R\, +\,  A_m^2) \  - \ \frac{M^2}{3} \ \chi^2 \ +  \ \frac{9}{2 M^2} \ ({\cal D} \cdot A)^2    \ , \label{dual13}
\end{equation}
and finally, eliminating $\chi$ via its algebraic field equation
\begin{equation}
\chi \ =\  \frac{3}{ {2} M^2} \ (R\, + \, A_m^2)\ ,  \label{dual14}
\end{equation}
one reaches the bosonic terms of the higher--derivative supergravity Lagrangian.

The final redefinition
\be
A_m \ \to \ \sqrt{\frac{2}{3}} \ A_m
\ee
recasts this Lagrangian in the notation of \cite{fv},
\begin{equation}
{\cal L} \ = \ \frac{1}{2}\ \left(R\, + \, \frac{2}{3}\ A_m^2\right)  \ + \ \frac{3}{4 M^2} \ \left(R\,+ \, \frac{2}{3}\ A_m^2\right)^2 \  +  \ \frac{3}{M^2} \ ({\cal D} \cdot A)^2
\ , \label{dual15}
\end{equation}
and the linearized higher--derivative terms then reproduce precisely the combination
\be
\frac{3}{4\, M^2} \ \big[ R^2 \ - \ 4\, A^\rho\, \partial_\rho \, \partial^{\,\mu} A_\mu \big] \ ,
\ee
in agreement with \cite{fgv}.

The bosonic Lagrangian (\ref{dual15}) propagates one scalar and one pseudoscalar degree of freedom, in addition to gravity. The scalar degree of freedom draws its origin, as is well known, from the $R^2$ term, which is manifest in our construction, while the need for the pseudoscalar one was pointed out long ago in \cite{fgv}. The field equation for $A_m$ following from eq.~\eqref{dual15} can be turned into a Klein-Gordon equation for ${\cal D} \cdot A$, since
\begin{equation}
\partial_m \, ({\cal D} \cdot A) \ - \ \frac{M^2}{9} \ A_m \ = \ 0 \qquad \rightarrow \qquad \Box \ ({\cal D} \cdot A) \ - \
\frac{M^2}{9} \ ({\cal D} \cdot A)\ =  \ 0 \ . \label{dual16}
\end{equation}
As a result, there is indeed one (pseudo)scalar degree of freedom whose mass, $\frac{M^2}{9}$, coincides with the mass of the dual axion $a$.

Let us conclude by stressing the dual gravitational interpretation of the constraint (\ref{dual3}), which translates into the component relations
\begin{eqnarray}
&& {\overline u} \ \equiv \  S \, + \, i \, P \ = \  M \ \frac{GG}{2 F_X} \ , \label{dual171} \\
&& \gamma^{mn} \, {\cal D}_m \, \psi_n \ = \  M \ G \ , \label{dual172} \\
&& - \frac{1}{2}\  R \ - \ \frac{1}{3}\  A_m^2 \ + \ i \, {\cal D}^m A_m \ - \ \frac{1}{3} \ u \, {\overline u} \ = \  M \ F_X \ ,
\end{eqnarray}
and thus links the (pseudo--)scalar auxiliary fields to the goldstino. Notice that in
this dual formulation the goldstino is determined by the gauge--invariant expression in eq.~\eqref{dual172}. All in all, the off--shell formulation is crucial for the consistency of the theory, and indeed the spacetime curvature $R$ is not fixed in any way by the constraint \eqref{dual4}, but is dynamically determined by the expectation value $\langle F_X \rangle$ of the auxiliary field $F_X$. Notice also that eq.~\eqref{dual171} implies that $u$ is nilpotent, a fact that we used in deriving the Lagrangian of eq.~\eqref{dual15}.

\vskip 24pt

\section*{Acknowledgments}

\vskip 10pt

It is a pleasure to acknowledge discussions with D.~Francia and A.~van Proeyen.
This work was supported in part by the ERC Advanced Grants n.~226455 (SUPERFIELDS) and n.~226371 (MassTeV), by Scuola Normale Superiore and by INFN (I.S.\ ST\&FI). The authors would like to thank the CPhT -- \'Ecole Polytechnique and the CERN Th-Ph Department for the kind hospitality extended to them while this work was in progress.

\end{document}

\end